%% file: wt.tex
\documentclass[12pt,a4paper]{article}
\usepackage{t1enc}
\usepackage[dvips,final]{graphicx} 
\usepackage{tabularx} 
\usepackage{amsfonts}
\usepackage{amsthm}

\def\du{\unskip\smash{\lower 1.4ex \hbox{\char34}}\kern-.2ex}
\def\hu{\kern-.2ex\hbox{\char92}}

\newcommand{\bdis}{\begin{displaymath}}
\newcommand{\edis}{\end{displaymath}}
\newcommand{\be}{\begin{equation}}
\newcommand{\ee}{\end{equation}}
\newcommand{\ot}{\otimes}
\newcommand{\pd}{\partial}

\begin{document}
\baselineskip=7.6mm
\newpage

\pagenumbering{roman} 

\title{How to Introduce A World Time to A Part of Two Dimensional de Sitter
Manifold}
\author{Michal Demetrian\footnote{{\it demetrian@sophia.dtp.fmph.uniba.sk}}
\qquad \\
{\it Dep. Theor. Phys., Commenius University} \\
{\it Mlynsk\' a Dolina, 842 48, Bratislava, Slovak Republic}}
\maketitle

\abstract{We define  1+1 dimensional de Sitter manifold in this
paper and we consider various coordinate systems on it. Some
interesting aspects of the general theory of relativity are
demonstrated using the transformations between considered
coordinate systems. The problem might be of interest for physics
and mathematics students as well as for physics teachers.}

\section{Introduction}
At first we stress the importance of $3+1$ version of de Sitter
manifold in relativistic cosmology (inflation) and quantum field
theory in curved spacetime. For more details see for example the
book of Birrell and Davies \cite{bd}, more information about the
(cosmological) inflation can be found in \cite{li} or in recent
papers concerning the inflation. We suppose that the reader of
this text is familiar with modern differential geometry. If not
see for example the introductory pages of books \cite{ch},
or \cite{mf}.  Now let us start with definition: \\
{\bf Def.:} 2D de Sitter manifold ($DS^2$) is defined by the
constraint
 \be x^2+y^2-t^2=a^2 \qquad (a>0) \quad ,
  \ee
as the submanifold of 3D Minkowski space $M^{2+1}_{[t,x,y]}$.
Metric structure on $M^{2+1}_{[t,x,y]}$ is given by the flat
metric tensor\footnote{we note, that the same as in eq. (2) can be
expressed using the infinitesimal interval as follows:
$ds^2=dx^2+dy^2-dt^2$, this notation is usual in many books
concerning both the special and the general theories of
relativity}
 \be \eta_3=dx\ot dx+dy\ot dy-dt\ot dt \quad .
 \ee

\hspace{1cm}
\begin{figure}[h]
\includegraphics[width=14cm, height=6cm]{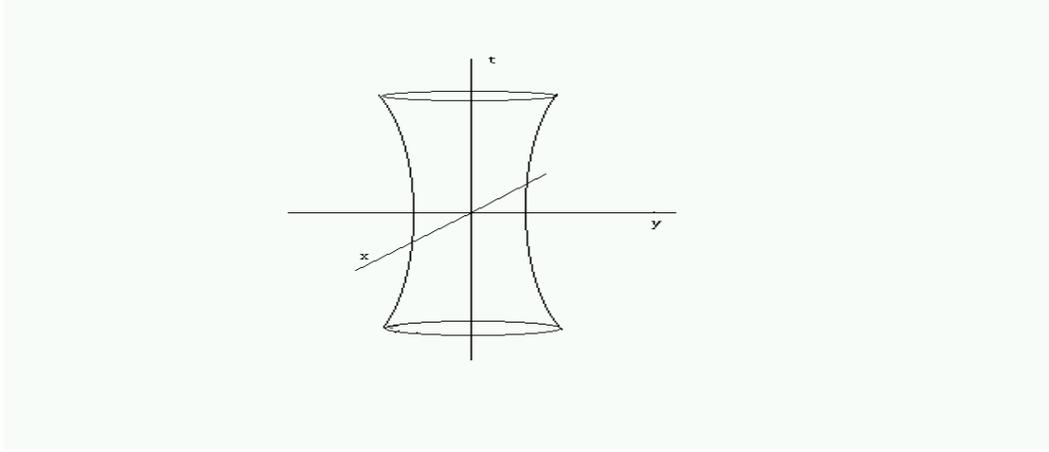}
\caption{$DS^2$ as the submanifold of $M^{2+1}$}
\label{fig.1.}
\end{figure}

So, if we introduce new coordinates ($r,\phi$) to describe $DS^2$
by the following equations
 \be
 \begin{array}{ccc}
 x & = & r\cos(\phi) \quad , \\
 y & = & r\sin(\phi) \quad ,
 \end{array}
 \qquad
 \begin{array}{c}
 r \in [a,\infty) \\
 \phi \in [0,2\pi) \quad ,
 \end{array}
 \ee
then we can write the induced metric tensor $g$ on $DS^2$ in the form:
 \be
 g=\frac{a^2}{a^2-r^2}dr\ot dr+r^2d\phi\ot d\phi \quad .
 \ee
Looking on the space-time character of vectors $\pd_r$ and
$\pd_\phi$ (i.e. evaluating $g(\pd_r,\pd_r)$ and
$g(\pd_\phi,\pd_\phi) )$ we shall see that the coordinates $\phi$
and $r$ correspond to space coordinate and time coordinate
respectively. The cut of $DS^2$ by the surface (see fig.1.)
$t=const$ is a circle parallel with the $xy$ plane, the radius of
the circle  is equal to $r=\sqrt{a^2+t^2}$. Such a cut represents
the ordinary space (or space), volume of which is equal to
 \be
 Vol_1=2\pi r=2\pi\sqrt{a^2+t^2} \quad .
 \ee
Another set of convenient coordinates ($\theta, \phi$) is given by
the transformation:
 \be r=a\cosh(\theta) \quad , \qquad \theta
 \in R \quad ,
 \ee
and the $g$ tensor has a form:
 \be g=a^2\left[ -d\theta \ot
 d\theta +\cosh^2(\theta)d\phi \ot d\phi \right] \quad .
 \ee
The last formula expresses that $DS^2$ is a two dimensional
version of the closed Robertson - Walker inflating Universe. The
coordinate $\theta$ plays the role of time and the scale parameter
(the radius of the space) evolution is given by
 \bdis
 a\cosh(\theta) \quad .
 \edis

\section{Killing's vector field on $DS^2$}

The isometric (continous) transformations of any smooth manifold
(especially $DS^2$) are generated by some vector field - say $V$.
We think by "are generated"  that the flow of $V$ is an isometric
transformation of the manifold under consideration. Such a vector
field $V$ is called the Killing's vector (vector field). The
condition for $V$ to be the Killing's vector is well known (the
Killing's equation) - see e.g. \cite{bd}:
 \be \mathcal{L}_Vg=0
 \quad ,
 \ee
where $\mathcal{L}_V$ is the Lie derivative with respect to $V$.
Let us denote $x^0=\theta$ and $x^1=\phi$ . Then the Killings'
equations for $DS^2$ may be written as follows
 \be
 V^kg_{ij,k}+V^k_{,i}g_{kj}+V^k_{,j}g_{ik} =0 \qquad (i,j,k\in\{
 0,1\}) \quad ,
 \ee
where $F_{,i}=\partial_{x_i}F$, or one can write the same in the
$(\theta,\phi)$ coordinates
 \be
 \begin{array}{ccc}
 V^1_{,1} & = & 0 \quad , \\
 \sinh(\theta)V^1+\cosh(\theta)V^2_{,2} & = & 0 \quad , \\
 \cosh^2(\theta)V^2_{,1}-V^1_{,2} & = & 0 \quad .
 \end{array}
 \ee
It is very simple to find the general solution of the system (10). The
result is
 \be
 V\equiv V^1\pd_1+V^2\pd_2=A\cos(\phi -\phi_0)\pd_{\theta}-A\tanh(\theta)
 \sin(\phi -\phi_0)\pd_{\phi}+B\pd_{\phi} ,
 \ee
with $A,B$ and $\phi_0$ being real constants. In the old
coordinates $(r,\phi)$ we have
 \bdis V=A\cos(\phi
 -\phi_0)\sqrt{r^2-a^2}\pd_r-A\sin(\phi -\phi_0)\frac{
 \sqrt{r^2-a^2}}{r}\pd_{\phi}+B\pd_{\phi} \quad .
 \edis

Now, we are going to separate $V$ into two parts and we shall
discuss their meaning: \\
(i.) Let us take
 \bdis V^{rot}=B\pd_{\phi} \quad .
 \edis
It is clear that the flow of
$V^{rot}$ makes the rotation of $DS^2$ with the velocity $B$
around the $t$ axis. (If
$B=0$ then we have the identity transformation.)
The space-time character of
$V^{rot}$ is given by its length
 \be
 g(V^{rot},V^{rot})=B^2g(\pd_{\phi},\pd_{\phi})=B^2r^2>0 \quad .
 \ee
So $V^{rot}$ is a space-like vector field and the integral curves of this
field are not worldlines of any real object. \\
(ii.) The rest of $V$ is
 \bdis
 \tilde{V} =A\sqrt{r^2-a^2}\left( \cos(\phi -\phi_0)\pd_r-\frac{1}{r}\sin(
 \phi -\phi_0)\pd_{\phi}\right) \quad .
 \edis
$DS^2$ is invariant under the flow of $B\pd_{\phi}$ (i.e. under
the rotations around the $t$ axis as was said in (i.)). So we can
put $\phi_0 =0$ without lost of generality. In this case we have
 \bdis
 \tilde{V}(\phi_0 =0)=A\sqrt{r^2-a^2}\pd_x \equiv \tilde{V}_x
 \quad .
 \edis
$\tilde{V}_x$ represents the (nonuniform) translation along the
$x$ axis projected on $DS^2$. (In the case of $\phi_0\not= 0$
$\tilde{V}$ represents analogical translation, but along some
axis, which is rotated relatively to the $x$ axis in the $xy$
plane of the background Minkowski space.) Let us look at the space
- time character of $\tilde{V}_x$
 \be
 g(\tilde{V}_x,\tilde{V}_x)=A^2(-a^2+y^2) \quad .
 \ee
We can state that the vector field $\tilde{V}_x$ is: \\
 - timelike in the part of $DS^2$, in which $|y|<a$ \\
 - spacelike in the part of $DS^2$, in which $|y|>a$ \\
The part of the space in which the inequality $|y|<a$ holds,
consists of two disjoint arcs - see fig.2. The $\phi$ angle, which
is defined on fig.2., is given at the moment $t$ as follows:
 \bdis
 \phi \in [-\alpha, \alpha] \qquad \mbox{if} \quad x>0 \quad ,
 \edis
 \bdis \phi \in [-\alpha + \pi, \alpha + \pi] \qquad
 \mbox{if} \quad x<0 \quad ,
 \edis
where
 \be \alpha = \arcsin\left( \frac{a}{r} \right) =
 \arcsin\left( \frac{a}{ \sqrt{a^2+t^2}}\right) \quad .
 \ee

\input{wt1.tex}
\input{wt2.tex}

\end{document}

%% file: wt1.tex
\section{Motion of light in $DS^2$}
A smooth curve (worldline)
$\gamma \subset DS^2$ is isotropic (or lightlike) if and only
if
 \be
 g(\dot{\gamma} ,\dot{\gamma})=0 \quad .
 \ee
A condition equivalent to this one is the following
 \be
 \frac{d\theta}{\cosh(\theta)}=\pm d\phi \quad .
 \ee
Performing the integration we get the equation of the general
isotropic line in $DS^2$ in the form
 \be \pm\phi =c_\pm
 +2\arctan(e^{\theta}) \quad .
 \ee
This solution can be written in other coordinates as
 \bdis \pm
 \phi =c_\pm + 2\arctan\left( \frac{r}{a}+\sqrt{ \left( \frac{r}{a}
 \right)^2-1} \right)= c_\pm +2\arctan \left( \sqrt{1+\left(
 \frac{t}{a} \right)^2}+\frac{t}{a} \right) .
 \edis
The space in $DS^2$ is one dimensional, the space coordinate from
the doublet $(t, \phi)$ is $\phi$ and $\phi \in [0,2\pi)$. The
difference
 \bdis \delta\phi=\phi(t_2)-\phi(t_1) \quad ,
 \edis
computed with respect to (17), describes the space translation of
the light during the time  interval between $t_1$ and $t_2$. If
two signals were emitted at the moment $t=0$ from the same point,
the first one in the $+\phi$ direction and the second in the
$-\phi$ direction, then the two points, that this two signals
reach at the time $t>0$, represent the particle horizon relatively
to the starting point of the signal emission. \\
Let us consider the two antiparallel signals emitted from the
point $(t=0, \phi=\pi/2) \in DS^2$ (see fig.2.). The constant $c$
(see (16)) is equal to $0$, for the signal emitted in $+\phi$
direction, and it is equal to $\pi$, for the signal emitted in the
$-\phi$ direction. The signals positions may be characterized by
the $\beta$ angle
 \bdis \beta(t) = \phi_+(t)-\phi_+(0) =
 2\arctan\left( \frac{t}{a}+\sqrt{\frac{t^2}{a^2}+1}\right)
 -\frac{\pi}{2} \quad .
 \edis
One trivially shows that
 \bdis
 \beta(t)=\frac{\pi}{2}-\alpha(t) \quad .
 \edis

 \begin{figure}[hc]
 \includegraphics[width=10cm, height=8cm]{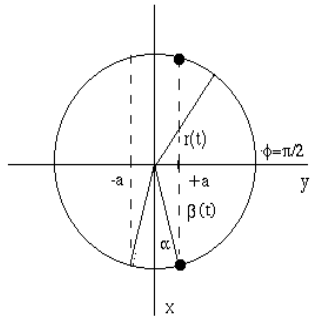}
 \caption{It shows the ordinary space and the particle horizon position
 (the two marked points) at the moment of time $t$}
 \label{fig.2.}
 \end{figure}

\section{World time}
The Killing's vector field $\tilde{V}_x$ is timelike in the part
of $DS^2$, in which the inequality $|y|<a$ is satisfied. It is
possible to introduce there local coordinates so that the
coefficients of the metric tensor do not depend on the time. One
speaks about the world time. We note, that this part of $DS^2$ is
assigned (fig.2.) by the $\alpha$ angle given by (14) or as well
as by $\beta$ angle and the
physical meaning of these angles was discussed in the previous sections. \\
In the area, which is now under consideration, we have
 \be
 x^2-t^2=a^2-y^2\equiv Y^2>0 \quad .
 \ee
So we can introduce the new coordinates $(Y,T)$ by the following relations
 \be
 \begin{array}{ccc}
 x & = & Y\cosh(T)\quad , \\
 t & = & Y\sinh(T) \quad ,
 \end{array}
 \qquad
 \begin{array}{ccc}
 Y & \in & [-a,a] \quad , \\
 T & \in & R \quad .
 \end{array}
 \ee
The metric tensor has the form
 \be g=\frac{a^2}{a^2-Y^2}dY\ot
 dY-Y^2dT\ot dT \quad .
 \ee
We see that we have presented a part of $DS^2$ as a static
spacetime and
that the $T$ coordinate plays the role of the world time. \\
What is the volume of the space (as a function of time) for an
observer measuring the space-time distances according to the
formula (20)? The space part of the metric tensor (19) is given by
 \bdis \hat{g}=\sqrt{|\det(g)|}dY=\frac{a}{\sqrt{a^2-Y^2}}dY \quad
 ,
 \edis
so we get the volume in question as follows
 \be
 Vol_2=a\int_{-a}^{+a}\frac{dY}{\sqrt{a^2-Y^2}}=\pi a \quad .
 \ee
We note that there are two (for all $t\not= 0$) disjoint parts of
$DS^2$, in which the world time can be introduced. The volume of
both of them is given by the previous formula.

%% file: wt2.tex
\section{Relativity of space infinity}
Now let us consider the part of $DS^2$, in which the inequality
$|y|>a$ holds.We can write
 \be x^2-t^2=a^2-y^2\equiv -Z^2<0 \quad,
 \ee
an we can make an analogical transformation of the coordinates as
in the previous section
 \be
 \begin{array}{ccc}
 x & = & Z\sinh(\rho)\quad ,  \\
 t & = & Z\cosh(\rho) \quad ,
 \end{array}
 \qquad
 \begin{array}{ccc}
 Z & \in & R \quad , \\
 \rho & \in & R \quad .
 \end{array}
 \ee
Tensor $g$ expressed in $(Y,\rho)$ coordinates becomes
 \be
 g=-\frac{a^2}{a^2+Z^2}dZ\ot dZ+Z^2d\rho\ot d\rho \quad .
 \ee
The last expression for $g$ suggests us to make one more
transformation
 \be
 Z=a\sinh(\tau)\quad , \qquad \tau\in R \quad ,
 \ee
and to express $g$ in terms of the pair $(\tau, \rho)$
 \be
 g=a^2\left[ -d\tau\ot d\tau +\sinh^2(\tau)d\rho\ot d\rho \right]
 \quad .
 \ee
It is clear that the $\tau$ coordinate plays the role of time now
and that we have just presented this part of $DS^2$ as the two
dimensional version of the open Robertson-Walker inflating
Universe with the scale parameter evolution given by
 \bdis
 a\sinh(\tau) \quad .
 \edis
The following formula
 \bdis \hat{g}=
 \sqrt{|\det(g)|}d\rho=a\sinh(\tau)d\rho \quad .
 \edis
gives the metrics on the hyperplane (the curve) $\tau=const.$. It
implies that the volume of space is given at the moment $\tau$ by
 \be
 Vol_3= a\sinh(\tau)\int_{-\infty}^{\infty}d\rho \quad .
 \ee
We can see that this volume is infinite for all $\tau \not= 0$.

\section{Conclusion}

We have started from the de Sitter manifold, which is a special
case of the closed Robertson-Walker Universe. Then we have
presented, using the suitable coordinate transformation (19), well
defined part of it as the static Universe, i.e. we have introduced
on that part of $DS^2$ the world time. One can discover on this
example, that the determination of the arrow of time, using the
expansion of the Universe, might be problematic. Thereafter we
used other coordinate transformations - eqs. (23) and (25) - and
we presented the part of $DS^2$ as the open Robertson-Walker
Universe. The motivation and the interpretation for these
coordinate transformations is given by the analysis of the
Killing's vector field od $DS^2$ and the motion of light in
$DS^2$, that we have done in the second and third sections. The
idea to present a part of closed inflating\footnote{by inflating
we think, that the scale parameter $a$ grows exponentially for
$t>0$} Universe as an open inflating Universe has a great use in
the cosmology, especially in the scenarios of evolution of the
very early Universe. From the physical point of view the inflating
Universe (as $DS^2$) is an Universe filled by the homogenous
configuration of a scalar field with the positive energy density.
This idea was presented in the work \cite{cl}. It leads to the
possibility of "the quantum creation of an open Universe" - see
for example \cite{li} or the short paper \cite{lin}.

%% file: wt.bbl
\begin{thebibliography}{10}
\bibitem{bd}
N.D.Birrell, P.C.W.Davies: Quantum Fields in Curved Space, Camb.
Univ. Press, 1982;
\bibitem{li}
A.D.Linde: Particle Physics and Inflationary Cosmology, Harwood,
Chur, Switzerland, 1990;
\bibitem{ch}
S.Chandrasekhar: The Mathematical Theory of Black Holes, Oxford
Univ. Press, Oxford, 1982;
\bibitem{mf}
M.Fecko: Differential Geometry for Physicists, to be published,
Bratislava 2002 (in slovak only);
\bibitem{cl}
S. Coleman, F. de Luccia: Phys. Rev. D {\bf 21}, 1980;
\bibitem{lin}
A.D. Linde: Phys. Rev. D {\bf 59}, 1999;
\end{thebibliography}
